\documentclass[10pt]{article}

\begin{document}

\title{Kaluza-Klein solitons reexamined}
\author{J. Ponce de Leon\thanks{E-mail: jpdel@ltp.upr.clu.edu, jpdel1@hotmail.com}\\ Laboratory of Theoretical Physics, Department of Physics\\ 
University of Puerto Rico, P.O. Box 23343, San Juan, \\ PR 00931, USA} 
\date{January  2007}

\maketitle

\begin{abstract}

In $(4 + 1)$ gravity the assumption that  the five-dimensional metric is independent of the  fifth coordinate authorizes the extra dimension to be either spacelike or timelike. As a  consequence of this,  the time coordinate and the extra coordinate are interchangeable, which in turn allows the conception of different scenarios in $4D$ from a single solution in $5D$. In this paper, we make a thorough investigation of all possible $4D$ scenarios, associated with this interchange, for the well-known Kramer-Gross-Perry-Davidson-Owen set of solutions. We show  that there are {\it three} families of solutions with very distinct geometrical and physical properties. They correspond to different sets of  values of the parameters which characterize the solutions in $5D$.  
The solutions of physical interest are identified on the basis  of physical requirements on the induced-matter in $4D$. 
We find that only one  family  satisfies these requirements; the other two violate the positivity of mass-energy density.  The ``physical" solutions possess a lightlike singularity which coincides with the horizon. The Schwarzschild black string solution as well as the zero moment dipole solution of Gross and Perry are obtained in different limits. These are analyzed in the context of Lake's geometrical approach. 
We demonstrate that the parameters of the  solutions in $5D$ are not free, as previously considered. Instead, they are  totally determined by measurements in $4D$. Namely, by the surface gravitational potential of the astrophysical phenomena, like the Sun or other stars, modeled in Kaluza-Klein theory. This is an important result which may help in observations for an experimental/observational test  of the theory.

\end{abstract}

\medskip



\newpage
\section{Introduction}

In four-dimensional general relativity, Birkhoff's theorem establishes that the Schwarzschild metric is the  only solution of the field equations\footnote{Our conventions are: c = G = 1; Greek indeces run over four-dimensions $\mu, \nu = 0, 1, 2, 3$; upper case Latin letters symbolize Kaluza-Klein indices $A, B = 0, 1, 2, 3, 4, ...$. The signature of the metric is $(1, -1, -1, -1, \pm 1).$}  $R_{\mu \nu} = 0$, with spherical symmetry. In more than four dimensions, in Kaluza-Klein theories,  this theorem is no longer valid: there are a number of solutions to the field equations $R_{AB} = 0$, with spherical three-space.

However, a  milder version of Birkhoff's theorem is true in Kaluza-Klein. Namely,   
there is only {\it one} family of  spherically symmetric exact solutions of the field equations $R_{AB} = 0$ which are asymptotically flat, static and independent of the ``extra" coordinates\footnote{The inclusion of spin changes this simple picture: Emparan and Reall \cite{Emparan1} have shown that, apart from the  black hole solutions of Myers and Perry \cite{Myers}, there are five-dimensional rotating black ring solutions, with the same values of mass and spin. }. In five-dimensions, in the form given by Davidson and Owen \cite{Davidson Owen}, they are described by the line element 
 
\begin{equation}
\label{Original Davidson and Owen solution}
dS^2 = \left(\frac{ar - 1}{ar + 1}\right)^{2\sigma k}dt^2 - \frac{1}{a^4r^4}\frac{(ar + 1)^{2[\sigma(k - 1) + 1]}}{(ar - 1)^{2[\sigma(k - 1) - 1]}}[dr^2 + r^2d\Omega^2]
- \left(\frac{ar + 1}{ar -1}\right)^{2\sigma}dy^2,
\end{equation}
where $d\Omega^2 = (d\theta^2 + \sin^2\theta d\phi^2)$; $(t, r, \theta, \phi)$ are the usual coordinates for a spacetime with spherically symmetric spatial sections; $y$ denotes the extra coordinate; $a$ is a constant with dimensions of $L^{- 1}$,  and $\sigma$ along with $k$ are parameters that obey the constraint 
\begin{equation}
\label{constraint on sigma and k}
\sigma^2(k^2 - k + 1) = 1.
\end{equation}
The above set of solutions has been rediscovered in different forms by Kramer \cite{Kramer} and, although in another context, by  Gross and Perry \cite{Gross Perry}. A particular case, in curvature coordinates, was given  by Chatterjee \cite{Chatterjee}  and more recently by  Millward \cite{Millward}.  They form a subset of the ``generalized Weyl solutions" of Emparan and Reall \cite{Emparan2} and are widely studied in the literature from different physical approaches. In particular they play a central role in the discussion of  
many important observational problems, which include the classical tests of relativity, as well as the geodesic precession of a gyroscope and possible departures from the equivalence principle \cite{Wesson book}. 

We note that in (\ref{Original Davidson and Owen solution}) the extra coordinate is spacelike. However, this is not a requirement of the field equations. Indeed, a closer examination of the field equations $R_{AB} = 0$, for solutions which are independent of the ``extra" coordinates, reveals that the large extra dimension $y$ can be either spacelike or timelike \cite{JpdeLgr-qc/0512067}. Thus, for generality, instead of (\ref{Original Davidson and Owen solution}) we should consider

\begin{equation}
\label{Davidson and Owen solution}
dS^2 = \left(\frac{ar - 1}{ar + 1}\right)^{2\sigma k}dt^2 - \frac{1}{a^4r^4}\frac{(ar + 1)^{2[\sigma(k - 1) + 1]}}{(ar - 1)^{2[\sigma(k - 1) - 1]}}[dr^2 + r^2d\Omega^2]
\pm \left(\frac{ar + 1}{ar -1}\right)^{2\sigma}dy^2.
\end{equation}

In this paper we study a number of aspects  that arise from the fact that by interchanging the roles of $t$ and $y$ in (\ref{Davidson and Owen solution}), and keeping the freedom for the signature of the extra dimension,  we generate the line element  

\begin{equation}
\label{Davidson and Owen solution with new parameterization}
d\bar{S}^2 = \left(\frac{ar + 1}{ar -1}\right)^{2\sigma}dt^2  - \frac{1}{a^4r^4}\frac{(ar + 1)^{2[\sigma(k - 1) + 1]}}{(ar - 1)^{2[\sigma(k - 1) - 1]}}[dr^2 + r^2d\Omega^2]
\pm  \left(\frac{ar - 1}{ar + 1}\right)^{2\sigma k}dy^2,
\end{equation}
which also satisfies the $5D$ field equations in vacuum. The issue is that 
metrics (\ref{Davidson and Owen solution}) and (\ref{Davidson and Owen solution with new parameterization}) seem to produce different interpretations on four-dimensional spacetime sections orthogonal to the extra dimension. In particular, the $5D$ analogue of the $4D$ Schwarzschild metric\footnote{This metric is known as the Schwarzschild black string.}, in isotropic coordinates
\begin{equation}
\label{Schw. limit}
dS_{Schw}^2 = \left(\frac{1 - M/2r}{1 + M/2r}\right)^2dt^2 - \left(1 + \frac{M}{2r}\right)^4[dr^2 + r^2d\Omega^2]
\pm dy^2,
\end{equation}
is recovered, for the same  central mass $M = 2/a$,  in different limits. Namely, $k \rightarrow \infty$ and $\sigma k \rightarrow 1$ for (\ref{Davidson and Owen solution}),  while $k = 0$ and $\sigma = - 1$ for (\ref{Davidson and Owen solution with new parameterization}).  

We note that in principle
\begin{equation}
\label{range of k}
- \infty  < k < + \infty,
\end{equation} 
and, as a consequence of the constraint (\ref{constraint on sigma and k}), $\sigma^2$ has a maximum, namely $\sigma^2 = 4/3$, at  $k = 1/2$. Therefore,
\begin{equation}
\label{range of sigma}
- \frac{2}{\sqrt{3}} \leq \sigma \leq \frac{2}{\sqrt{3}}.
\end{equation}

The  first goal of this work is  to elucidate the link  between the four-dimensional interpretation of metrics (\ref{Davidson and Owen solution}) and (\ref{Davidson and Owen solution with new parameterization}) 
(on hypersurfaces $y =$ constant) and their parameters $k$ and $\sigma$.  In this regard a number of questions arise.  For example, what is the ``appropriate"  range of the parameters $k$ and $\sigma$?; what is the physical meaning of $k$?. In this paper we discuss these questions in the context of the induced-matter interpretation. We recall that, in this context the curvature in $5D$ induces effective matter in $4D$, and the metric  (\ref{Davidson and Owen solution}) can be interpreted as describing extended spherical objects called solitons. 

In section $2$, in order to identify the solutions of physical interest we impose  physical requirements on the induced-matter. We find that these conditions demand $k > 0$ and $\sigma > 0$ for metric  (\ref{Davidson and Owen solution}), while $k > 0$ and $\sigma < 0$ for metric (\ref{Davidson and Owen solution with new parameterization}). We show that, although the parameters $\sigma$ and $k$ are {\it not} independent, the transformation from metric (\ref{Davidson and Owen solution}) to (\ref{Davidson and Owen solution with new parameterization}), and vice-versa, corresponds to the  {\it simultaneous} change $k \rightarrow 1/k$, $\sigma \rightarrow - \sigma$, or 
\begin{equation}
\label{corresponding change}
\left(\sigma k\right) {\leftrightarrow} - \sigma,
\end{equation}
which leaves $\sigma(k - 1)$ and $\sigma^2 k$ invariant. 
   
In a recent  paper, Lake  \cite{Lake} examines the properties of the Kramer-Gross-Perry-Davidson-Owen solutions in a purely geometrical way. The solutions are classified on the basis of the Weyl invariant, the geometrical mass and the character of the singularity. Therefore his results in $5D$ hold  for any physical approach in $4D$.

The  second goal of this paper is to find out how the physics in $4D$ is subordinated  to the general geometrical properties in $5D$.  Since  the properties of the effective-matter depend on $k$ and $\sigma$, we need to relate these parameters to those of Lake \cite{Lake}. In section $3$, we provide a complete analysis of the metrics (\ref{Davidson and Owen solution}) and (\ref{Davidson and Owen solution with new parameterization}) in the context of the geometrical approach.  Our analysis is very similar, but not identical, to Lake's and leads to somewhat distinct results in $4D$. For example, metrics (\ref{Davidson and Owen solution}) and (\ref{Davidson and Owen solution with new parameterization})  require $ar \geq 1$, which in terms of the coordinate $h$ used by Lake corresponds to $h \in (1, \infty)$; the region $h \in (0, 1)$ considered in \cite{Lake} is excluded here because it is not asymptotically flat\footnote{As a consequence,  
the geometrical symmetry between quadrant $2$ for $h \in (1, \infty)$ and quadrant $1$ for $h \in (0, 1)$ is broken.}. 
We will see that this ``lack" of symmetry results in two families of solutions with very different physical properties in $4D$.

The third goal here is to understand the physical meaning of the parameter $k$. In section $4$, we demonstrate that $k$ is completely determined by the {\it degree of compactification} of the soliton. Thus, $k$ is   neither a universal constant nor a free parameter,  but varies from soliton to soliton. This feature has been overlooked in our previous work \cite{previous work} and other subsequent studies \cite{Wesson book}, \cite{Agnese}, \cite{Sajko}.

In the Appendix we examine the more general case where the metric in $4D$ is taken to be conformal to the metric induced from $5D$ on four-dimensional hypersurfaces orthogonal to the extra dimension.  We find the same kind of solutions as in the induced-matter approach, but with a different parameterization.

\section{The induced-matter approach} 

The aim of this section is to compare and contrast the four-dimensional interpretation of metrics (\ref{Davidson and Owen solution}) and (\ref{Davidson and Owen solution with new parameterization}). In order to facilitate the presentation, let us restate some concepts that are essential  in our discussion. 

In five-dimensional  models, our spacetime is identified with some  $4D$ hypersurface $y = $ constant, which is orthogonal to the extra dimension. Therefore, for a given line element in $5D$  
\begin{equation}
\label{general metric}
dS^2 = g_{\mu\nu}(x^{\rho}, y)dx^{\mu}dx^{\nu} + \epsilon \Phi^2(x^{\rho}, y)dy^2,\;\;\;\epsilon = \pm 1,
\end{equation}
the corresponding metric in $4D$ is just $g_{\mu\nu}$. Such an identification is  a standard\footnote{For an alternative approach, which  reproduces our $4D$ spacetime on a {\it dynamical} hypersurface,   see Refs. \cite{JpdeLgr-qc/0511067} and \cite{JPdeLgr-qc/0511150}.}  technique in the induced-matter approach as well as in brane-world models. However, it is worth to mention the approach where the  metric in $4D$ is conformal to the metric induced on $y$ = constant hypersurfaces. We will examine this approach in the Appendix. 

In $4D$, the effective energy-momentum tensor $T_{\mu\nu}$  is obtained from the $4 + 1$ dimensional  reduction of the field equations in $5D$. In terms of the metric, it is given by \cite{JPdeL'sEMT}  
\begin {eqnarray}
\label{EMT in STM}
8 \pi T_{\alpha\beta} =  &-& \frac{\epsilon}{2\Phi^2}\left[\frac{\stackrel{\ast}{\Phi} \stackrel{\ast}{g}_{\alpha \beta}}{\Phi} - \stackrel{\ast \ast}{g}_{\alpha \beta} + g^{\lambda\mu}\stackrel{\ast}{g}_{\alpha\lambda}\stackrel{\ast}{g}_{\beta\mu} - \frac{1}{2}g^{\mu\nu}\stackrel{\ast}{g}_{\mu\nu}\stackrel{\ast}{g}_{\alpha\beta} + \frac{1}{4}g_{\alpha\beta}\left(\stackrel{\ast}{g}^{\mu\nu}\stackrel{\ast}{g}_{\mu\nu} + (g^{\mu\nu}\stackrel{\ast}{g}_{\mu\nu})^2\right)\right] \nonumber \\
 &+& \frac{\Phi_{\alpha;\beta}}{\Phi}, 
\end{eqnarray}
where $\stackrel{\ast}{f} \equiv \partial f/\partial y$. For the case where the $5D$ metric is independent of $y$, the effective matter is {\it not} affected by the signature of the extra dimension and $T_{\mu\nu}$ reduces to 
\begin {equation}
\label{EMT for the solutions under consideration}
8 \pi T_{\mu\nu} =   \frac{\Phi_{\mu;\nu}}{\Phi},
\end{equation}
with $g^{\mu\nu}\Phi_{\mu;\nu} = 0$, which follows from $R_{44} = 0$. What this means is that, in this case the effective matter in $4D$ is radiation like.\footnote{It is sometimes called ``black" or ``Weyl" radiation, because in this case $T_{\mu\nu} = - \epsilon E_{\mu\nu}$, where $E_{\mu\nu}$ represents the spacetime projection of the five-dimensional Weyl tensor, which is traceless.}

In the case under consideration the $5D$ metric has the form 
\begin{equation}
\label{general static metric}
dS^2 = e^{\nu(r)}dt^2 - e^{\lambda(r)}[dr^2 + r^2d\Omega^2]
+ \epsilon \Phi^2(r)dy^2.
\end{equation}
Then, from $g^{\mu\nu}\Phi_{\mu;\nu} = 0$, it follows that
\begin{equation}
\label{equation for Phi}
\Phi'' = - \Phi'\left(\frac{\nu' + \lambda'}{2} + \frac{2}{r}\right),
\end{equation}
where primes denote derivatives with respect to $r$. Using this expression, the explicit form of the induced energy-momentum tensor (\ref{EMT for the solutions under consideration}) can be written as 
\begin{eqnarray}
\label{EMT in terms of Phi}
8\pi T^0_0 &=& - \frac{e^{- \lambda}\Phi'\nu'}{2 \Phi}
\nonumber, \\ 
8\pi T^1_1 &=& e^{- \lambda}\left(\frac{\nu'}{2} + \lambda' + \frac{2}{r}\right)\frac{\Phi'}{\Phi}\nonumber, \\
8\pi T^2_2 &=& 8\pi T^3_3 = - e^{- \lambda}\left(\frac{1}{r} + \frac{\lambda'}{2}\right)\frac{\Phi'}{\Phi}.
\end{eqnarray}
\subsection{Physical radius}

For metrics (\ref{Davidson and Owen solution}) and (\ref{Davidson and Owen solution with new parameterization}) the ``physical" radius $R$ of the sphere with coordinate radius $r$ is given by
\begin{equation}
\label{physical radius}
R(r) = \frac{(ar + 1)^{\sigma(k - 1) + 1}}{a^2 r(ar - 1)^{\sigma(k - 1) - 1}}.
\end{equation}
We note that for  $(k > 0$, $\sigma > 0)$, $(k > 0$, $\sigma < 0)$, $(k < 0$, $\sigma > 0)$, and  $(k = 0, \sigma = 1)$, the center of the sphere $R = 0$ corresponds to  $ar = 1$ and $R$ increases monotonically with the increase or $r$, i.e., $(dR/dr) > 0$. 

However, there is no origin if we choose either $(k < 0$,  $\sigma < 0)$ or $(k = 0, \sigma = -1)$. Indeed, for this choice $(dR/dr)$ changes sign at

\begin{equation}
\label{minimum}
ar_{min} = \frac{- \sqrt{k^2 + |k| + 1} + |k| + 1 + \sqrt{|k|}}{\sqrt{k^2 + |k| + 1}} + 1.
\end{equation}
Thus, for $(k < 0$,  $\sigma < 0)$,  $R \rightarrow \infty$ as $ar \rightarrow 1$ and $ar \rightarrow \infty$. The radius $R$ has a minimum at the value of $ar$ given by (\ref{minimum}). For $(k = 0, \sigma = - 1)$, $R \rightarrow \infty$ as $ar \rightarrow 0$ and $ar \rightarrow \infty$. We note that $R$ is not well defined\footnote{One could think that for $\sigma(k - 1) - 1 = 2n$, where $n$ is some integer number, one could properly define $R$, for $ar < 1$,  as $R = (ar + 1)^{2(n + 1)}/[a^2r(1 - ar)^{2n}]$. In this case $\sigma(k - 1) = 2n + 1$ and substituting into (\ref{constraint on sigma and k}) we obtain a quadratic  equation for $k$, namely, $4(n^2 + n) k^2 - (4n^2 + 4n - 1)k + 4(n^2 + n) = 0$. However, for an arbitrary $n$ this equation has no real solution. There are only two real solutions: for $n = -1$ and $n = 0$. They correspond to  the special cases $(k = 0, \sigma = 1)$  and $(k = 0, \sigma = - 1)$ considered above.} for $ar < 1$. Therefore, in what follows we will consider $ar \geq 1$ everywhere.

\paragraph{Sign of $k$:} From (\ref{EMT in terms of Phi}) we obtain 
 \begin{equation}
\label{positivity of mass-energy density}
8 \pi T_{0}^{0} = \frac{4 a^6 \sigma^2 k r^4}{(ar + 1)^4(ar - 1)^4}\left(\frac{ar - 1}{ar + 1}\right)^{2\sigma(k - 1)}.
\end{equation}
for metrics (\ref{Davidson and Owen solution}) and (\ref{Davidson and Owen solution with new parameterization}).  We note that this is invariant under transformation (\ref{corresponding change}). Consequently, the positivity of mass-energy density requires $k > 0$ for both metrics, which in turn assures that $R = 0$ at $ar = 1$ and $dR/dr > 0$ everywhere. 

\subsection{Gravitational mass}

In $4D$, the  gravitational mass inside a $3D$ volume $V_{3}$ is given by the Tolman-Whittaker formula, viz.,
\begin{equation}
\label{standard gravitational mass}
M_{g}(r) = \int{(T^0_0 - T^1_1 - T^2_2 - T^3_3)}\sqrt{- g_{4}}dV_{3}.
\end{equation}
Using (\ref{EMT in terms of Phi}) we obtain
\begin{equation}
\label{grav. mass in terms of Phi}
M_{g}(r) = - \frac{1}{2}\int_{1/a}^{r}{e^{(\nu + \lambda)/2}}\frac{\nu' \Phi'}{\Phi}r^2dr.
\end{equation}
For the metric  (\ref{Davidson and Owen solution}), after straightforward calculation we get 
\begin{equation}
M_{g}(r) = \frac{(2\sigma k)}{a}\left(\frac{ar - 1}{ar + 1}\right)^{\sigma },
\end{equation}
while for the metric\footnote{In what follows  quantities, as $M_{g}, \sigma$ and others, corresponding to metric (\ref{Davidson and Owen solution with new parameterization})  will be denoted with a bar over them, i.e., ${\bar{M}}_{g}, \bar{\sigma}$, etc.} (\ref{Davidson and Owen solution with new parameterization}) 
\begin{equation}
\label{grav mass for DO}
\bar{M}_{g}(r) = \frac{(- 2\sigma)}{a}\left(\frac{ar + 1}{ar - 1}\right)^{\sigma k}.
\end{equation}
Clearly, the interchange $\sigma k \leftrightarrow - \sigma$ transforms  $M_{g}(r) \leftrightarrow \bar{M}_{g}(r)$. 
 
\paragraph{Positivity of gravitational mass: Sign of $\sigma$.} Since $a$ is related to the Schwarzschild mass we take $a > 0$ everywhere. Therefore, the positivity of the gravitational mass $M_{g}$, for metric (\ref{Davidson and Owen solution}) requires $\sigma > 0$, i.e., 
\begin{equation}
\label{positive sigma}
\sigma = + \frac{1}{\sqrt{k^2 - k + 1}}. 
\end{equation}
On the other hand, for the metric (\ref{Davidson and Owen solution with new parameterization}) the positivity of  $\bar{M}_{g}$ requires $\sigma < 0$, i.e.,

\begin{equation}
\label{negative sigma}
\bar{\sigma} = - \frac{1}{\sqrt{k^2 - k + 1}}.
\end{equation}
In summary, the gravitational mass becomes 
\begin{equation}
\label{grav. mass for sol. 1}
M_{g}(r) = \frac{2k}{a\sqrt{k^2 - k + 1}}\left(\frac{ar - 1}{ar + 1}\right)^{1/\sqrt{k^2 -k + 1}}, 
\end{equation}
and 
\begin{equation}
\bar{M}_{g}(r) = \frac{2}{a\sqrt{k^2 - k + 1}}\left(\frac{ar - 1}{ar + 1}\right)^{k/\sqrt{k^2 -k + 1}}, 
\end{equation}
for (\ref{Davidson and Owen solution}) and (\ref{Davidson and Owen solution with new parameterization}), respectively.

\subsection{Possible scenarios in $4D$}
Thus, in the full range of $k$ and $\sigma$, there are solutions   
with distinct physical properties. 
Namely, the original Davidson-Owen family of solutions (\ref{Davidson and Owen solution}) contain four different scenarios. These are\footnote{The identification used here for the distinct solutions  is similar to the one used by Lake \cite{Lake}.}:
\begin{eqnarray}
\label{families in DO}
1: (k < 0, \sigma > 0) \leftrightarrow (\rho < 0, M_{g} < 0),\nonumber \\ 
2: (k < 0, \sigma < 0) \leftrightarrow (\rho < 0, M_{g} > 0), \nonumber \\
3: (k > 0, \sigma < 0) \leftrightarrow (\rho > 0, M_{g} < 0), \nonumber \\
4: (k > 0, \sigma > 0) \leftrightarrow (\rho > 0, M_{g} > 0).
\end{eqnarray}
Under the transformation $t \rightarrow y$ we get  the metric (\ref{Davidson and Owen solution with new parameterization}), which allows the following scenarios
\begin{eqnarray}
\label{families in new DO}
\bar{1}: (k < 0, \sigma > 0) \leftrightarrow (\rho < 0, M_{g} < 0),\nonumber \\ 
\bar{2}: (k < 0, \sigma < 0) \leftrightarrow (\rho < 0, M_{g} > 0),\nonumber \\
\bar{3}: (k > 0, \sigma < 0) \leftrightarrow (\rho > 0, M_{g} > 0),\nonumber \\
\bar{4}: (k > 0, \sigma > 0) \leftrightarrow (\rho > 0, M_{g} < 0).
\end{eqnarray}
All these families, except for $2$ and $\bar{2}$,  have a center at $ar = 1$,  where $M_{g}$ = 0,  as well as  $dR/dr >0$ everywhere. Solutions $2$ and $\bar{2}$ have no center and $M_{g} \rightarrow \infty$ for $ar \rightarrow 1$. In summary: 

\begin{enumerate}
\item The physical properties of the first two solutions are {\it invariant}
 under the transformation $t \leftrightarrow y$, i.e., $1\leftrightarrow \bar{1}$ and $2\leftrightarrow \bar{2}$. 

\item The other two solutions show interchange symmetry, i.e., $3 \leftrightarrow \bar{4}$ and  $4 \leftrightarrow \bar{3}$.

\item Solutions with $\rho > 0$ and $M_{g} < 0$, after the transformation $t\leftrightarrow y $ become $\rho > 0$ and $M_{g} > 0$. Namely, $3 \rightarrow \bar{3}$ and $\bar{4} \rightarrow 4$.

\end{enumerate}

\subsection{Effective matter for solutions $4$ and $\bar{3}$}

These satisfy the requirements on the induced-matter in $4D$. In Appendix A, we show that $T^1_1 \neq T^2_2$ is a general feature of solitons in theories where the metric in $4D$ is conformal  to the metric induced  induced on $y$ = constant hypersurfaces. Therefore, the generic approach is to describe the soliton matter as an anisotropic fluid with an effective energy-momentum tensor of the form\footnote{Under certain conditions, a single anisotropic fluid can be modeled as a multicomponent fluid \cite{Letelier}-\cite{Comment}.}
\begin{equation}
\label{EMT for anisotropic matter}
T_{\mu\nu} = (\rho + p_{\perp})u_{\mu}u_{\nu} - p_{\perp}g_{\mu\nu} + (p_{r} - p_{\perp})\chi_{\mu}\chi_{\nu}, 
\end{equation}
where $u^{\mu}$ is the four-velocity; $\chi^{\mu}$ is a unit spacelike vector orthogonal to $u^{\mu}$; $\rho$ is the energy density; $p_{r}$ is the pressure in the direction of $\chi_{\mu}$, and $p_{\perp}$ is the pressure on the two-space orthogonal to $\chi_{\mu}$. If we choose $u^{\mu} = \delta^{\mu}_{0}e^{- \nu/2}$ and $\chi^{\mu} = \delta^{\mu}_{1}e^{- \lambda/2}$, then $T^0_{0} = \rho$, $T^1_{1} = - p_{r}$ and $T^2_{2} = - p_{\perp}$. Consequently, the equation of state becomes 
\begin{equation}
\label{equation of state for soliton matter}
\rho = p_{r} + 2p_{\perp},
\end{equation}
which shows that the matter has the nature of radiation. 

\paragraph{Effective matter for solution $4$:}
Collecting results, the evaluation of the induced-matter quantities (\ref{EMT in terms of Phi}) for the solution $4$ in (\ref{families in DO}) yields 
\bigskip

\begin{equation}
\label{matter density for the DO solution}
8 \pi \rho= \frac{4 a^6  k r^4}{(k^2 - k + 1)(ar + 1)^4(ar - 1)^4}\left(\frac{ar - 1}{ar + 1}\right)^{2(k - 1)/\sqrt{k^2 - k + 1}}\;\;\;\;\;\;\;\;\;\;\;\;\;\;\;\;\;\;,
\end{equation}

\bigskip

\begin{equation}
8 \pi p_{r} = \frac{4 a^5 r^3[ar(2 - k) + (a^2r^2 + 1)\sqrt{k^2 - k + 1}]}{(k^2 - k + 1)(ar + 1)^4(ar - 1)^4}\left(\frac{ar - 1}{ar + 1}\right)^{2(k - 1)/\sqrt{k^2 - k + 1}}\;\;\;,
\end{equation}

\bigskip

\begin{equation}
\label{tangential p for DO solution}
8 \pi p_{\perp} = \frac{2 a^5r^3[2ar(k - 1) - (a^2r^2 + 1)\sqrt{k^2 - k + 1})]}{(k^2 - k + 1)(ar + 1)^4(ar - 1)^4}\left(\frac{ar - 1}{ar + 1}\right)^{2(k - 1)/\sqrt{k^2 - k + 1}}.
\end{equation}
We note that $\rho = p_r = p_{\perp} = 0$ for $k \rightarrow \infty$, as expected.
\paragraph{Effective matter for solution $\bar{3}$:}

Similarly, for the solution $\bar{3}$ in (\ref{families in new DO}) we find
\begin{equation}
\label{rho for the DO sol in the new parameterization}
8 \pi \bar{\rho} = \frac{4 a^6 k r^4}{(k^2 - k + 1)(ar + 1)^4(ar - 1)^4}\left(\frac{ar - 1}{ar + 1}\right)^{2(1 - k)/\sqrt{k^2 - k + 1}}\;\;\;\;\;\;\;\;\;\;\;\;\;\;\;\;\;\;\;,
\end{equation}

\bigskip

\begin{equation}
8 \pi \bar{p}_{r} = 
\frac{4ka^5 r^3[ar(2k -1) + (a^2r^2 + 1)\sqrt{k^2 - k + 1}]}
{(k^2 - k + 1)(ar + 1)^4(ar - 1)^4}\left(\frac{ar - 1}{ar + 1}\right)^{2(1 - k)/\sqrt{k^2 - k + 1}}, 
\end{equation}

\bigskip

\begin{equation}
\label{p perp for the DO sol in new parameterization}
8 \pi \bar{p}_{\perp} = 
\frac{2ka^5 r^3[2ar(1 - k) - (a^2r^2 + 1)\sqrt{k^2 - k + 1}]}{(k^2 - k + 1)(ar + 1)^4(ar - 1)^4}\left(\frac{ar - 1}{ar + 1}\right)^{2(1 - k)/\sqrt{k^2 - k + 1}}.
\end{equation}
Here $\bar{\rho} = \bar{p}_r = \bar{p}_{\perp} = 0$ for $k = 0.$ 

\bigskip
Clearly, 
\begin{equation}
(\rho, \;\;p_{r}, \;\;p_{\perp}) {\longleftrightarrow} (\bar{\rho}, \;\;\bar{p}_{r}, \;\;\bar{p}_{\perp})
\end{equation}
 for 
\begin{equation}
k {\longleftrightarrow} 1/k.
\end{equation}
Both distributions are identical for $k = 1$, but for any other $k$ they are very different\footnote{For  $k = 1$ the solution takes a particular simple form. It was rediscovered by  Chatterjee \cite{Chatterjee}.}.

Finally, we mention that for possible astrophysical applications of solitons, it is crucial   to note that Kaluza-Klein solitons are more massive than the Schwarzschild one. Indeed, we find 
\begin{equation}
\label{confinement of Mg}
\frac{2}{a} \leq M_{g}(\infty) \leq \frac{4}{a \sqrt{3}}.
\end{equation}
This is an  interesting result which advocates  for solitons as candidates for dark matter \cite{Wesson book}. 

\section{The geometrical approach}

In a recent paper, Lake examined the properties of the Kramer-Gross-Perry-Davidson-Owen solutions in a purely geometrical way \cite{Lake}. He  classified  the solutions on the basis of the Weyl invariant, the nakedness and geometrical mass\footnote{The gravitational mass defined in (\ref{standard gravitational mass}) is not equivalent to the geometrical mass, which is  defined via the sectional curvature of the two-sphere \cite{Lake}.} of their associated singularities. The natural question to ask is how the properties of induced matter in $4D$ are related, or subordinated, to the geometrical ones in $5D$.  In order to facilitate the discussion, in this section we use the codification of the solutions used by Lake.

\subsection{Lake's parameterization}

In Lake's work the solutions are described in terms of the parameters $\alpha$ and $\delta$, in such a way that  the Davidson-Owen line element (\ref{Davidson and Owen solution}) is recovered by changing $\delta \rightarrow - 2\sigma k$ and $\alpha \rightarrow 2\sigma$.
Under the transformation
\begin{equation}
\label{Lake's transformation}
\sigma = \frac{\alpha}{2}, \;\;\; k = - \frac{\delta}{\alpha},
\end{equation}
 the constraint (\ref{constraint on sigma and k}) becomes
\begin{equation}
\label{Lake's parameterization}
\alpha^2 + \delta^2 + \alpha \delta = 4,
\end{equation}
which in the $(\alpha, \delta)$ plane represents an ellipse\footnote{A similar parameterization was considered by Lim, Overduin and Wesson \cite{LOW}.}. Conversely,  setting $\alpha = 2\sigma$ and  $\delta = - 2\sigma k$ we recover (\ref{constraint on sigma and k}).

As we have discussed in section $2.1$, the physical radius in Davidson-Owen solutions, in the coordinates of (\ref{Davidson and Owen solution}),   is well defined in the region  $ar \geq 1$ only. This corresponds to $h \in(1, \infty)$ in Lake's notation\footnote{The region $h \in (0, 1)$ is excluded from our discussion because those solutions are not asymptotically flat. As a consequence, we loose the  symmetry between quadrant $2$ for $h \in (1, \infty)$ and quadrant $1$ for $h \in (0, 1)$. But the symmetry between solutions in quadrants $3$ and $4$ is not affected. See Table $1$ in Ref. \cite{Lake}.}. In this region there are {\it three} ``regular" solutions which, in the attached figure, correspond to  quadrants $1, 3$ and $4$. For these solutions $R = 0$ and $M_{g} = 0$ at $ar = 1$, besides $dR/dr > 0$. Quadrant $2$ solutions are singular in the sense that there is no origin and $M_{g} \rightarrow \infty$ in the limit $ar \rightarrow 1$, which now corresponds to $R \rightarrow \infty$. Also, there are four ``exceptional" solutions, namely ${a} = (2, 0),  {b} = (0, 2),  {c} = (- 2, 0)$ and ${d} = (0, - 2)$. 

Regular solutions in quadrants $1$ and $4$ have positive $\sigma$, viz., $\sigma = + 1/\sqrt{k^2 - k + 1}$. So, in our approach,  they are described by the original Davidson-Owen line element (\ref{Davidson and Owen solution}).  

In quadrants $1$ and $4$ the parameter $k$ increases clockwise, along the ellipse, from $k = - \infty$ at the exceptional solution $b$, to $k = 0$ at $a$ and $k = + \infty $ at $d$. Thus, $k < 0$ in quadrant $1$ and $k > 0$ in quadrant $4$.  From (\ref{positivity of mass-energy density}) it follows that in quadrant $1$ the energy condition  $\rho > 0$ is violated. Meanwhile, in quadrant $4$  the effective matter distribution, which is  given by(\ref{matter density for the DO solution})-(\ref{tangential p for DO solution}), satisfies the physical conditions $\rho > 0$, $M_{g} >0$ and possesses an origin $R = 0$ at $ar = 1$.

The line element corresponding to the exceptional solution $a = (2, 0)$ is obtained from the metric (\ref{Davidson and Owen solution}) for $k = 0$ and $\sigma = 1$, 
\begin{equation}
\label{solution a}
dS^2_{a(k = 0, \sigma = 1)} = dt^2 - \left(1 - \frac{1}{ar}\right)^4[dr^2 + r^2d\Omega^2]
\pm \left(\frac{1 + 1/ar}{1 -1/ar}\right)^{2}dy^2.
\end{equation} 

Quadrant $2$ singular solutions and quadrant $3$ regular solutions have negative $\sigma$, viz., $\sigma = - 1/\sqrt{k^2 - k + 1}$. So, in our approach,  they are described by the line element (\ref{Davidson and Owen solution with new parameterization}).  In these quadrants, the parameter $k$ also increases clockwise, along the ellipse. It goes  from $k = - \infty$ at $d$, to $k = 0$ at $c$ and $k = + \infty $ at $b$. Thus, $k < 0$ in quadrant $2$ but $k > 0$ in quadrant $3$.  Therefore, $\rho < 0$ in $2$ but in  $3$  the effective matter distribution, which is  given by (\ref{rho for the DO sol in the new parameterization})-(\ref{p perp for the DO sol in new parameterization}), satisfies the physical conditions $\rho > 0$, $M_{g} > 0$ and possesses an origin $R = 0$ at $ar = 1$.

The line element corresponding to the exceptional solution $c = (- 2, 0)$ is obtained from the metric (\ref{Davidson and Owen solution with new parameterization}) for $k = 0$ and $\sigma = - 1$. Namely,
\begin{equation}
\label{Schw solution from DO with new parameterization}
d\bar{S}^2_{c(k = 0, \sigma = -1)} = \left(\frac{1 - 1/ar}{1 + 1/ar}\right)^{2}dt^2  - \left(1 + \frac{1}{ar}\right)^4[dr^2 + r^2d\Omega^2]
\pm  dy^2,
\end{equation}
which is the $5D$ analogue of the $4D$ Schwarzschild metric in isotropic coordinates  with $a = 2/M$. 

The point  $b = (0, 2)$ is attained from quadrant $1$ (say $b_{1}$ from metric (\ref{Davidson and Owen solution})),  for $\sigma = 0$, $k = - \infty$ and from quadrant $3$ (say $b_{3}$ from metric (\ref{Davidson and Owen solution with new parameterization})),  for $ \sigma = 0, k = + \infty$. Therefore, there are two limiting metrics 
\begin{equation}
\label{solution b for dS}
 dS^2_{b_{1}} = dS^2_{b(k = - \infty, \sigma = 0)} = \left(\frac{1 + 1/ar}{1 - 1/ar}\right)^2dt^2 - \left(1 - \frac{1}{ar}\right)^4[dr^2 + r^2d\Omega^2]
  \pm dy^2.
\end{equation} 
\begin{equation}
\label{solution b for dS bar}
 d\bar{S}^2_{b_{3}} =   d\bar{S}^2_{b(k = + \infty, \sigma = 0)} = dt^2  - \left(1 - \frac{1}{ar}\right)^4[dr^2 + r^2d\Omega^2]
\pm  \left(\frac{1 + 1/ar}{1 - 1/ar}\right)^2dy^2,
\end{equation}

The point $d = (0, - 2)$ is attained from quadrant $4$ (say $d_{4}$ from metric (\ref{Davidson and Owen solution})), for $\sigma = 0$, $k = + \infty$ and from quadrant $2$ (say $d_{2}$ from metric (\ref{Davidson and Owen solution with new parameterization})) for $ \sigma = 0, k = - \infty$, viz.,
\begin{equation}
\label{solution d for dS}
dS^2_{d_{4}} = dS^2_{d(k = \infty, \sigma = 0)} = \left(\frac{1 - 1/ar}{1 + 1/ar}\right)^{2}dt^2  - \left(1 + \frac{1}{ar}\right)^4[dr^2 + r^2d\Omega^2]
\pm  dy^2  
\end{equation} 
\begin{equation}
\label{sol d for dS bar}
d\bar{S}^2_{d_{2}} = d\bar{S}^2_{d(k = - \infty, \sigma = 0)} = dt^2  - \left(1 + \frac{1}{ar}\right)^4[dr^2 + r^2d\Omega^2]
\pm  \left(\frac{1 - 1/ar}{1 + 1/ar}\right)^2dy^2.
\end{equation}
Clearly, by changing $t \leftrightarrow y$ we convert $dS^2_{b} \leftrightarrow d\bar{S}^2_{b}$ $(b_{1} \leftrightarrow b_{3})$ and $dS^2_{d} \leftrightarrow d\bar{S}^2_{d}$ $(d_{2} \leftrightarrow d_{4})$. No such connection exists between solutions $a$ and $c$. The metrics (\ref{solution b for dS}) and (\ref{solution d for dS}) represent the Schwarzschild black string in $5D$ with $M = - 2/a$ and $M = 2/a$, respectively. On the other hand, solutions (\ref{solution a}), (\ref{solution b for dS bar}) and (\ref{sol d for dS bar}) represent the   zero dipole moment soliton of Gross and Perry \cite{Gross Perry}. 

The top line $\alpha + \delta = 2/\sqrt{3}$ connects the solution $e = (4/\sqrt{3}, - 2/\sqrt{3})$, for which $k = 1/2$ and $\sigma = 2/\sqrt{3}$,  with the solution $\bar{e} = (- 2/\sqrt{3}, 4/\sqrt{3})$, for which $k = 2$ and $\sigma = - 1/\sqrt{3}$.

The bottom line $\alpha + \delta = - 2/\sqrt{3}$ connects the solution $f = (2/\sqrt{3}, - 4/\sqrt{3})$, for which $k = 2$ and $\sigma = 1/\sqrt{3}$,  with the solution $\bar{f} = (- 4/\sqrt{3}, 2/\sqrt{3})$, for which $k = 1/2$ and $\sigma = - 2/\sqrt{3}$.

Solutions of quadrant $2$ have been interpreted as describing wormholes by Agnese {\it et al} \cite{Agnese}.
They possess positive gravitational mass,   but violate the weak energy condition $\rho > 0$. The solution given recently by Millward is located in quadrant $1$ and corresponds to the particular choice $k = - 1$, $\sigma = 1/\sqrt{3}$ or $\alpha = \delta = 2/\sqrt{3}$ (see \cite{Lake}). Therefore,  it exhibits negative gravitational mass.

\subsection{Interchange symmetry and physical equivalence}

The ellipse (\ref{Lake's parameterization}) is invariant under the change  $\alpha \rightarrow - \alpha$, $\delta \rightarrow - \delta$, which is equivalent to rotating the ellipse in $180^{\circ}$, in any direction. In terms of the Davidson-Owen parameters, this corresponds to the transformation 
\begin{equation}
\label{interchange symmetry}
\sigma = - \frac{\alpha}{2}, \;\;\; k =  - \frac{\delta}{\alpha},
\end{equation} 
It should be noted that setting  $(\alpha = - 2\sigma, \delta = 2 k \sigma)$ in Lake's solution,  we recover the line element (\ref{Davidson and Owen solution with new parameterization}) instead of (\ref{Davidson and Owen solution}). 

From (\ref{Lake's transformation}) and (\ref{interchange symmetry}), it follows that the interchange $(\alpha, \delta) \leftrightarrow (- \alpha, - \delta)$ is analogous to the choice of positive or negative $\sigma$, keeping the same $k$,  as in (\ref{positive sigma}) and (\ref{negative sigma}). Consequently, if we use  (\ref{interchange symmetry}) instead of (\ref{Lake's transformation}), then we obtain the quadrant interchanges $1 \leftrightarrow 2$ and $3 \leftrightarrow 4$ (say $\bar{e} \leftrightarrow f$ and $\bar{f} \leftrightarrow e$)  along with the point interchanges $a \leftrightarrow c$, i.e.,  $dS^2_{a}  \leftrightarrow d\bar{S}^2_{c}$. Regarding solutions $b$ and $d$ the invariance $\alpha \rightarrow - \alpha$, $\delta \rightarrow - \delta$ corresponds to $dS^2_{b}  \leftrightarrow d\bar{S}^2_{d}$ ($b_{1} \leftrightarrow d_{2}$) and $dS^2_{d}  \leftrightarrow d\bar{S}^2_{b}$ ($b_{3} \leftrightarrow d_{4}$).

However, it should be emphasized that this geometrical invariance is not accompanied by  a ``physical" equivalence. For example, it transforms the  black string (\ref{Schw solution from DO with new parameterization}) into the  zero dipole moment soliton (\ref{solution a}). In the induced-matter approach, the effective energy-momentum tensors corresponding to $\bar{e}$ and $f$ (as well as $\bar{f}$ and $e$) are {\it totally} different. 

\subsection{Singularities and $t \leftrightarrow y$}

We notice that  the invariance under $\alpha \leftrightarrow \delta$ is equivalent to (\ref{corresponding change}). This symmetry is {\it not} a consequence of  any rotation in the $(\alpha, \delta)$ plane\footnote{A rotation in $90^{\circ}$ would be $\alpha \rightarrow \delta, \delta\rightarrow - \alpha$}, but is a consequence of the interchange $t \leftrightarrow y$, which is allowed by the freedom of the signature of the extra dimension in Ricci flat $5D$ manifolds with spatial spherical symmetry and no-dependence of the extra dimension.

The singularity at $R = 0$ $(ar = 1)$, for   solutions in quadrants $3$ and $4$, corresponds to a lightlike singularity. The same kind of naked singularities, where the horizon coincides with the singularity,  are found in black hole solutions $(R_{AB} = 0)$ in other dimensions, for example in $d = 11$ supergravity \cite{Baal}.

Solutions $c$ and $b_{3}$ correspond, respectively,  to the Schwarzschild solution with a spacelike singularity and the zero dipole moment soliton with timelike singularity. Thus, in quadrant $3$ as $k$ goes from zero to infinity, the singularity changes from spacelike to lightlike and then to timelike. Similarly, in quadrant $4$ as $k$ goes from zero to infinity, the singularity changes from timelike at $a$ to lightlike and then to spacelike at $d_{4}$.

\section{Degree of compactification}

The soliton matter is distributed in the form of centrally concentrated clouds, {\it without} a solid surface. However, the matter density decreases as $\rho \sim 1/a^2r^4$ indicating that the matter is heavily concentrated near the origin. Therefore, it is always possible to define a sphere where most of the total mass is contained. 

Let us define $r_{\xi}$, which represents the coordinate radius of the sphere  containing the $\xi$-th part of the total gravitational matter of the soliton $(0 \leq \xi \leq 1)$. 

The total gravitational mass for the soliton described by metric (\ref{Davidson and Owen solution}) is $M_{g}(\infty) = 2k/(a\sqrt{k^2 - k +1})$, which is 
obtained from (\ref{grav. mass for sol. 1}) in the limit $ar \gg 1$. Therefore, we find
\begin{equation}
r_{\xi}(k) = \left(\frac{1 + \xi^{\sqrt{k^2 - k + 1}}}{1 - \xi^{\sqrt{k^2 - k + 1}}}\right)\frac{1}{a}.
\end{equation}
 The corresponding physical radius $R_{r_{\xi}} = r_{\xi}e^{\lambda(r_{\xi})/2}$ is
\begin{equation}
\label{physical radius}
R_{\xi}(k) = \frac{4}{a(1 - \xi^{2\sqrt{k^2 - k + 1}})\xi^{[(k - 1) - \sqrt{k^2 - k + 1}]}}
\end{equation}
We now define the ``surface" gravitational potential $\phi$ as
\begin{equation}
\phi_{\xi} = \frac{M_{\xi}}{R_{\xi}},
\end{equation}
which for the case under consideration becomes
\begin{equation}
\label{gravitational potential}
\phi_{\xi}(k) = \frac{1}{2}\frac{k}{\sqrt{k^2 - k + 1}}(1 - \xi^{2\sqrt{k^2 - k + 1}})\xi^{[k - \sqrt{k^2 - k +1}]}.
\end{equation}
This function has two important features. Namely, it is independent of parameter $a$, and is a monotonic function of $k$. Therefore, it gives a one-to-one connection between the surface gravitational potential $\phi$ and the soliton parameter $k$, which allows us to calculate this parameter for different astrophysical phenomena. 

\subsection{Evaluation of $k$}

In order to study observational implications of extra dimensions, and test possible deviations from general relativity, the Sun and other stars are modeled as Kaluza-Klein solitons \cite{Liu-Overduin}, \cite{Overduin}. Let us apply the above formulae to evaluate $k$ for the solar system.

 In this case, $\phi$ can be taken as  the surface gravitational potential of the Sun, which  is\footnote{Here  we use $M_{\odot} = 1.9891 \times 10^{30} kg$  and that the mean diameter of the Sun is about 
$1.392 \times 10^{6} km$ . Also, since $c = G = 1$, it follows that $1 kg = 7.41 \times 10^{- 31} km$ and $M_{\odot} = 1.474 km$ .} $\phi_{\odot} = 0.212 \times 10^{- 5}$. 
Although the Sun is a gaseous sphere, without  a sharp boundary, most of its mass is contained within the photosphere of (mean) radius $R_{\odot} = 0.696 \times 10^{6} km$. For the sake of argument, let us say that $99.9999 \%$ of the total mass  is enclosed there\footnote{This is within the accuracy allowed by the standard model for the interior of the Sun (no rotation, no diffusion) by Pierre Demarque and David Guenther in \cite{book}. For this selection, the remaining $10^{- 6}M_{\odot}$ part is distributed above the photosphere with very low density. If we use (\ref{matter density for the DO solution}), we find that $90\%$ of this part is located in a layer, whose thickness is $9$ times the solar radius, which is about  $(1/10)$-th of the mean distance between the Sun and Mercury.}
, which corresponds to  $\xi = 0.999999$.  Substituting these values into (\ref{gravitational potential}) we obtain 
\begin{equation}
k_{\odot} = 2.12.
\end{equation} 
In order to see whether this number makes sense we use it in (\ref{matter density for the DO solution}) to evaluate the energy density at the surface of the sun. We obtain $\rho(R_{\odot}) \approx 0.23 \times 10^{- 6} gr/cm^3$, which is very close to the average density of the photosphere, whose accepted value $\rho_{ph} = 0.22 \times 10^ {- 6} gr/cm^3$ is given in Ref.  \cite{book}.

If we model more compact astrophysical objects, like neutron stars, as Kaluza-Klein solitons, we find that  $k$ is significantly larger. For example, for the value of $\xi$ taken above and the  surface potentials
\begin{equation}
\phi = (10^{- 3},\;\;10^{- 2},\;\;10^{- 1},\;\;2\times 10^{- 1},\;\;3\times 10^{- 1}, \;\;4\times 10^{- 1}),
\end{equation}
from (\ref{gravitational potential}) we obtain
\begin{equation}
k \approx (10^3, \;10^4,\;1.1\times 10^5,\;2.5\times 10^5,\;4.5 \times 10^5,\;8 \times 10^5).
\end{equation}
Clearly, $k \rightarrow \infty$ for $\phi \rightarrow 1/2$, which corresponds to the Schwarzschild black hole. A similar result can be obtained from metric (\ref{Davidson and Owen solution with new parameterization}) with the corresponding change (\ref{corresponding change}). The main conclusion from this section is that the Kaluza-Klein parameter $k$ is not a free one, as previously considered. Instead, it  is totally determined by the surface gravitational potential of the astrophysical phenomena, like the Sun or other stars, modeled in Kaluza-Klein theory. 

\section{Summary and concluding remarks}

When the metric coefficients are independent of $y$, the extra dimension can be either spacelike or timelike, without affecting the effective matter distribution \cite{JpdeLgr-qc/0512067}. This is true for both, static and non-static metrics. In practice this means that $t$ and $y$ are interchangeable, which in turn allows us to generate different scenarios in $4D$. 

The crucial question is, how do we recover our $4$-dimensional world in higher-dimensional theories?. This is an unsettled question yet, but the popular wisdom is that we do this by going onto a hypersurface $y$ = constant.  With this assumption, the information about the fifth dimension is consolidated in the  nonlocal stresses induced in $4D$ from the Weyl curvature in $5D$.  As a consequence,  even in the absence of matter, the exterior of a spherical star, like our sun,  is {\it not} in general an empty Schwarzschild spacetime. We expect that  these stresses, contributing to the effective pressure, will influence  the conditions at the surface of a star \cite{Prep}.

In this work we have discussed in detail the relationship between the apparently different $4D$ scenarios, belonging to the well-known Kramers-Gross-Perry-Davidson-Owen family of solutions. 
In sections $2$ and $3$ we have provided a detailed study of the metrics (\ref{Davidson and Owen solution}) and (\ref{Davidson and Owen solution with new parameterization}) within the context of the induced-matter and the geometrical approaches. 

The solutions of quadrants $3$ and $4$ share the same geometrical properties in $5D$ (see Table $1$ in Ref. \cite{Lake}). We have found that their four-dimensional counterparts are equivalent modulo transformation $k \leftrightarrow 1/k$; the corresponding matter quantities  transform as $(\rho, \;\;p_{r}, \;\;p_{\perp}) {\longleftrightarrow} (\bar{\rho}, \;\;\bar{p}_{r}, \;\;\bar{p}_{\perp})$ for $k \leftrightarrow 1/k$. 
 Besides, as $k$ increases from zero to infinity, the singularity of quadrants $3$ solutions changes as: (spacelike $\rightarrow$ lightlight $\rightarrow$ timelike). For the same range of $k$, the  singularity of  quadrant $4$ solutions changes as: (timelike $\rightarrow$ lightlight $\rightarrow$ spacelike). 

The solutions of quadrants $1$ and $2$ are symmetrical in $5D$, and duplicate each other, iff we include in the discussion  the region $0 < ar < 1$  ($h \in (0, 1)$ in Lake's notation \cite{Lake}). However, when we restrict ourselves to solutions that are asymptotically flat, as we do here, then we  exclude the region $0 < ar < 1$, which breaks the  symmetry between them. The corresponding  models in $4D$ bear the imprint of this lack of symmetry.  Namely, the physical radius of quadrant solutions $2$ has a positive minimum at the value of $r$ given by (\ref{minimum}), while quadrant solutions $2$ have an origin $R = 0$ at $ar = 1$ and $dR/dr > 0$ everywhere.  In  the context of induced-matter approach,  this lack of symmetry yields solutions with positive gravitational mass for $(k < 0,  \sigma < 0)$, and  negative gravitational mass for $(k < 0, \sigma > 0)$.  

\medskip

In summary, the main conclusions from this paper are the following: 
\begin{enumerate}
\item In the range $ar \geq 1$ (or $h \in(1, \infty)$ in Lake's notation), the solutions in quadrants $1$, $2$ and  $3$ (or $4$) are all distinct from each other; they have different physical and geometrical properties. They epitomize the character and nature of the effective four-dimensional picture in the conformal approach (\ref{conformal approach}), where we find the same kind of solutions as in the induced-matter approach, but with a different parameterization. 

\item Results concerning the signs of $k$ and $\sigma$, and their relations to Lake's
parameters allow us to compare and contrast both approaches. In the geometrical approach, based on the nature of the singularities at the origin, in \cite{Lake} it is concluded that the physical solutions are those of quadrant $1$ and $2$.  
However, in the induced matter approach the solutions of physical interest are $3$ and $4$. Since the black string and the zero dipole moment solitons are unstable to any metric perturbation, one would expect them to  decay into quadrant solutions $3$ or $4$, because the weak energy condition is violated in $1$ and $2$.

 \item The singularity at $r = a$, in the physical region $3$ (or $4$), is lightlike and coincides with the horizon. 
\item Solutions with $\rho >0$ and $M_{g} < 0$, after the transformation $t\leftrightarrow y$ become identical to those with positive $\rho$ and $M_{g}$. This seems to be a general feature of solutions admitting a timelike extra dimension. 
\item The basic characteristic  of the effective soliton matter is that it  behaves like  an anisotropic fluid, which can be described by the energy-momentum tensor (\ref{EMT for anisotropic matter}). This is true   in the induced-matter approach, as well as in the conformal approach (\ref{conformal approach}). This is a general result which, as far as we know, has never been unveiled in the literature. 
\item The five-dimensional parameter $k$ can be evaluated by means of measurements in $4D$, namely by the ``surface" gravitational potential of the soliton. This is an important result which may help in observations for the experimental/observational test of the $5D$ theory.  
\end{enumerate}

\renewcommand{\theequation}{A-\arabic{equation}}
  % redefine the command that creates the equation no.
  \setcounter{equation}{0}  % reset counter 
  \section*{Appendix A: The  conformal approach}  % use *-form to suppress numbering

The prediction of  physical effects from extra dimensions, which can be measured in experiments or observations, requires a ``correct" identification of the physical or observable spacetime from the multidimensional one. Unfortunately, this is not an easy task \cite{JpdeLgr-qc/0512067}. 

Besides the approach discussed in section $2$, we would like to briefly discuss the approach where the effective metric in $4D$, say $g_{\mu\nu}^{eff}$, is conformal to the metric induced from $5D$ on four-dimensional hypersurfaces orthogonal to the extra dimension. 
The factorization  $g_{\mu\nu}^{eff} = \Phi g_{\mu\nu}$  has been considered as a standard Kaluza-Klein technique in $5D$ theories with a compact extra dimension\footnote{A similar technique   has been used in more that $5$-dimensions \cite{Dereli}} \cite{Davidson Owen}, \cite{Dolan-Duff}. Here, for generality,  we will examine 

\begin{equation}
\label{conformal approach}
 g_{\mu\nu}^{eff} = \Phi^N g_{\mu\nu},
\end{equation}
where $N$ is some constant. 
For $N = 1$ we have the original Davidson-Owen interpretation, and for $N = 0$ the  induced-matter interpretation discussed here. 

For future references we provide the non-vanishing components of the effective energy-momentum tensor. For the metric (\ref{Original Davidson and Owen solution}) these are 

\begin{equation}
8\pi T^{0}_{0} = \frac{a^6 r^4\sigma^2[4k - N^2  - 4N(k - 1)]}{(ar + 1)^4(ar - 1)^4}\left(\frac{ar - 1}{ar + 1}\right)^{2\sigma(k - 1 + N/2)},
\end{equation}

\begin{equation}
8 \pi T_{1}^{1} = - \frac{4 a^5 r^3\sigma^2 \left\{(1 -N)[ar(2 - k) + (a^2r^2 +1)/\sigma] + (3/4)N^2 ar\right\}}{(ar + 1)^4(ar - 1)^4} \left(\frac{ar - 1}{ar + 1}\right)^{2\sigma(k - 1 + N/2)}\;\;\;,
\end{equation}

\begin{equation}
8 \pi T_{2}^{2} = 8 \pi T_{3}^{3}  = - \frac{2 a^5r^3 \sigma^2\left\{2ar[k(1 - N) - 1] - (1 - N)(a^2r^2 + 1)/\sigma) + (1/2)N^2 ar\right\}}{(ar + 1)^4(ar - 1)^4}\left(\frac{ar - 1}{ar + 1}\right)^{2\sigma(k - 1 + N/2)}.
\end{equation}

\paragraph{Anisotropic pressures:}

From the above we get 
\begin{equation}
8 \pi (T_{2}^{2} - T_{1}^{1}) = - \frac{2 a^5r^3 \sigma^2\left\{[4k(1 - N) + 4N - 6 - N^2]ar + 3(N - 1)(a^2r^2 + 1)/\sigma\right\}}{(ar + 1)^4(ar - 1)^4}\left(\frac{ar - 1}{ar + 1}\right)^{2\sigma(k - 1 + N/2)},
\end{equation}
which shows that $T_{2}^{2} \neq  T_{1}^{1}$ for {\it any} value of $N$. Thus, the effective matter  behaves  as an anisotropic fluid, which can be described by the energy-momentum tensor (\ref{EMT for anisotropic matter}), not only in the induced-matter approach. In fact, we see that this is a general feature of the conformal approach (\ref{conformal approach}). This is an interesting result which, as far as we know, has never been exposed in the literature.   

\paragraph{Gravitational mass:} It can be verified that the gravitational mass, given by the Tolman-Whittaker formula (\ref{standard gravitational mass}), becomes 

\begin{equation}
M_{g}(r) =  \frac{(2k - N)\sigma}{a}\left(\frac{ar - 1}{ar +1}\right)^{\sigma(1 - N)}.
\end{equation}
Here, we require $(2k - N)\sigma > 0$ and ${\sigma(1 - N) > 0}$ in order to ensure the positivity of $M_{g}$, as well as, the condition $M_{g} = 0$ at $ar = 1$, respectively.  

\paragraph{Factorization with $N = 0$:} This case corresponds to the induced-matter and braneworld approaches where the metric of the physical spacetime is identified with the metric induced in $4D$. In this case the above expressions reduce to the ones in section $2$. 

\paragraph{Factorization with $N \neq 1$:}

This is the general case. Now  the condition $\rho > 0$ demands $k > [N(N - 4)/4(1 - N)]$ for $N < 1$, and $k < [N(N - 4)/4(1 - N)]$ for $N > 1$. Once again, there are four different families of solutions. They are equivalent to those in (\ref{families in DO}) for $N = 0$, but are displaced along the ellipse.  For example, for $N = 2$ the physical solutions ($\rho >  0$; $M_{g} > 0$, with a center at $ar = 1$ where $M_{g} = 0$)  move from quadrant $4$ to the region encompassing quadrant 2 and the lower half (bellow the dotted line) of quadrant 3, i.e., $(k < 1, \sigma < 0)$. The wormhole solutions of Agnese {\it et al} \cite{Agnese} are now parameterized by $(k > 1, \sigma > 0)$. So they move from quadrant $2$ to the lower half (bellow the dotted line) of quadrant $4$. 

\paragraph{Factorization with $N = 1$:}
This factorization was considered by Davidson and Owen \cite{Davidson Owen}. In this case the expressions for the effective energy-momentum tensor are specially simple. However, as far as we know, they are unreachable in the literature. Therefore, in view of their  importance, we provide them here.
  
For the five-dimensional metric (\ref{Original Davidson and Owen solution}), the effective spacetime is described by the line element 
\begin{equation}
ds^2 = \left(\frac{ar - 1}{ar  + 1}\right)^{2\varepsilon}dt^2  - \frac{1}{a^4r^4}\frac{(ar + 1)^{2[\varepsilon + 1]}}{(ar - 1)^{2[\varepsilon - 1]}}[dr^2 + r^2d\Omega^2],
\end{equation}
where $\varepsilon = \sigma(k - 1/2)$.

In this case 
\begin{equation}
8\pi \rho = \frac{4a^6 r^4(1 - \varepsilon^2)}{(ar + 1)^4(ar - 1)^4}\left(\frac{ar - 1}{ar + 1}\right)^{2\varepsilon}.
\end{equation}
The equations of state for the anisotropic matter are 
\begin{equation}
\label{equations of state}
p_{r} = \rho,\;\;\;p_{\perp} = - \rho.
\end{equation} 
If we introduce the concept of average pressure, $<p>$, as
\begin{equation}
\label{average pressure}
<p>  = - \frac{1}{3}(T_{1}^{1} + T_{2}^{2} + T_{3}^{3}),
\end{equation}
then, the effective matter satisfies the equation of state\footnote{In the induced-matter approach, from (\ref{equation of state for soliton matter}) it follows that $\rho = 3<p>$. However, we should  emphasize that this and (\ref{equation of state for soliton matter in the conformal approach}) are just  {\it formal} expressions, because the physical interpretation of an average pressure, in different ``directions", is not clear.} 
\begin{equation}
\label{equation of state for soliton matter in the conformal approach}
\rho = - 3<p>.
\end{equation}

The gravitational mass for $N = 1$ is constant throughout the space, viz.,
\begin{equation}
M_{g}(r) = \frac{2\varepsilon}{a}.
\end{equation}
The Schwarzschild limit corresponds to $\varepsilon = 1$, for which  we get $M_{g} = 2/a$, as expected. We note that the {\it only} conditions on the ``four-dimensional" parameter $\varepsilon$ come from the positivity of $\rho$ and $M_{g}$, which require
\begin{equation}
0 \leq \varepsilon \leq 1.
\end{equation}
Once we know its value, we can reconstruct the five-dimensional quantities $\sigma$ and $k$ as follows 
\begin{equation}
\sigma = \pm \frac{2\sqrt{1 - \varepsilon^2}}{\sqrt{3}},\;\;\;k = \frac{1}{2}\left(1 \pm \frac{\sqrt{3} \varepsilon}{\sqrt{1 - \varepsilon^2}}\right)
\end{equation}

Finally, we would like to mention that under the transformation $t \leftrightarrow y$, the above expressions should be changed according to (\ref{corresponding change}), i.e., $\sigma k \leftrightarrow - \sigma$ everywhere.

\end{document}